\documentclass[a4paper]{article}

\usepackage{icrc2013}
          
\title{The lateral shower age parameter as an estimator of chemical composition}

\shorttitle{A. Tapia \textit{et al.} The lateral shower age and chemical composition}

\authors{
A. Tapia$^{1}$,
D. Melo$^{1}$,
F. S\'anchez$^{1}$,
A. Sedoski Croce$^{1}$,
A. Etchegoyen$^{1}$,
J. M. Figueira$^{1}$,
R. F. Gamarra$^{1}$,
B. Garc\'ia$^{2}$,
N. Gonz\'alez$^{1}$,
M. Josebachuili$^{1}$,
D. Ravignani$^{1}$,
I. Sidelnik$^{1,3}$,
B. Wundheiler$^{1}$.
}

\afiliations{
$^1$ ITeDA (CNEA, CONICET, UNSAM), Buenos Aires, Argentina. \\
$^2$ ITeDA (CNEA, CONICET, UNSAM), Mendoza, Argentina. \\
\scriptsize{
$^3$ now at: Centro At\'omico Bariloche (CNEA y CONICET), Av. Bustillo 9500 (8400), San Carlos de Bariloche, R\'io Negro, Argentina. \\
}
}

\email{alex.tapia@iteda.cnea.gov.ar}

\abstract{We explore the feasibility of estimating primary cosmic ray composition at ultra high energies from the study of lateral age parameter of Extensive Air Showers (EAS) at ground level. Using different types of lateral distribution functions, we fit the particle density of simulated EAS to find the lateral age parameter. We discuss the chemical composition calculating the merit factor for each parameter distribution. The analysis considers three different primary particles (proton, iron and gamma), four different zenith angles (0$^{\circ}$, 15$^{\circ}$, 30$^{\circ}$ and 45$^{\circ}$) and three primary energies ($10^{17.25}$ eV, $10^{17.50}$ eV and $10^{17.75}$ eV).}

\keywords{Extensive air showers, lateral shower age parameter, chemical composition.}

\begin{document}
\maketitle

\section{Introduction}
The particle lateral distribution of extensive air showers (EAS) is the key quantity for cosmic ray ground observations, from which most shower observables are derived. An EAS is initiated by a high energy cosmic ray particle in the atmosphere, creating a multitude of secondary particles, which arrive nearly at the same time distributed over a large area perpendicular to the direction of the original particle. The disc of secondary particles may extend over several hundred meters from the shower axis, reaching its maximum density in the center of the disc, which is called the shower core. The density distribution of particles within the shower disc can be used to derive information on the primary particle. Due to the low rate of these events on the earth surface, EAS measurements on ground level are carried out using large arrays of individual detectors, which take samples of the shower disc at several locations \cite{bib:Roth, bib:kascade}.\\

One of the parameters commonly used to describe the form of the lateral density distribution is the \textit{lateral shower age parameter} (LSAP) in the Nishimura-Kamata-Greisen (NKG)-function \cite{bib:Nishimura, bib:Greisen}. The name LSAP expresses the relation between the lateral shape of the electron distribution and the height of the shower maximum. Due to the statistical nature of the shower development, the height of the shower maximum is subject to strong fluctuations. Showers, which have started shallower in the atmosphere are called old and they are characterised by a large LSAP value. Young showers have started deeper in the atmosphere, which corresponds to a smaller value of the LSAP. Apart from fluctuations, the height of the shower maximum depends on energy and mass of the shower initiating primary \cite{bib:Nagano}. Therefore, the LSAP is also sensitive to the mass of the primary \cite{bib:kascade}.\\

In the following sections, using the concept of LSAP, we will present the study of the chemical composition. First, we will show the LSAP calculated with fits of the simulated lateral density distribution using two types of functions: NKG \cite{bib:Nishimura, bib:Greisen} and Linsley \cite{bib:Linsley}. Secondly, we find the distributions of LSAP for each primary particle, zenith angle and energy used. Finally, by mean of the merit factor between the distributions we analyze the composition discrimination power of the LSAP.
\section{Monte Carlo simulation}
For this work we generated a library of extensive air showers using AIRES 2.8.4a \cite{Aires:02} based on the
hadronic model QGSJET-II-03 \cite{QGSJET:04,QGSJET:05}. We set a relative thinning of $10^{-6}$, a weight factor of $0.2$ and a ratio between the two weight factors (electromagnetic/hadronic) equal to 88.\\

We consider three types of primaries (proton, iron and gamma), four zenith angles (0$^{\circ}$, 15$^{\circ}$, 30$^{\circ}$ and 45$^{\circ}$) and three energies ($10^{17.25}$ eV, $10^{17.50}$ eV and $10^{17.75}$ eV).\\

For each energy, zenith angle and primary type a total of 600 showers were produced considering a uniform azimuthal distribution between 0$^{\circ}$ and 360$^{\circ}$.
Only gammas, electrons/positrons and muons with energies above 1.286 MeV, 264 keV and 55 MeV, respectively, have been taken into account.
\section{Lateral shower age}
The LSAP was introduced primarily to describe the development of the electromagnetic cascade. Nishimura-Kamata and Greisen found a function which relates the lateral distribution of shower particles with the shower age. That was called NKG function, which has the form:
\begin{equation}\label{eq:fNKG}
 \rho (R)=C \frac{N}{R_{0}^{2}}\left(\frac{R}{R_0}\right)^{s-2}\left(1+\frac{R}{R_0}\right)^{s-4.5},
\end{equation}
where $\rho(R)$ is the particle density at distance $R$, $N$ is the total number of shower secondaries, $C$ is the normalization constant, $R_0$ is the Moliere unit and $s$ is the shower age \cite{bib:Nishimura, bib:Greisen}.\\

Since the 70's, many authors have pointed out that NKG function with a single age parameter is not adequate to describe the lateral distribution at all distances, meaning that lateral shower age varies with the axial distance \cite{bib:Dai}. Linsley has proposed a double parameter function, characterized by $\alpha$ and $\eta$ \cite{bib:Linsley}, to make a correction of the NKG function, expressed as:
\begin{equation}\label{eq:fLins}
 \rho (R)=C(\alpha, \eta)\frac{N}{R_{0}^{2}}\left(\frac{R}{R_0}\right)^{-\alpha}\left(1+\frac{R}{R_0}\right)^{-(\eta-\alpha)}.
\end{equation}
The distribution is determined by $\alpha$ when $R$ approches $0$, and by $\eta$ when $R$ approches $\infty$.\\

Performing the first derivative of Eq. (\ref{eq:fNKG}), one can get:
\begin{equation}\label{eq:dNKG}
 \frac{d\ln\rho(r)}{d\ln r}=(s-2)+\frac{(s-4.5)r}{(1+r)}
\end{equation}
and from Linsley's function, Eq. (\ref{eq:fLins}):
\begin{equation}\label{eq:dLins}
 \frac{d\ln\rho(r)}{d\ln r}=-\frac{(\alpha + r\eta)}{(1+r)},
\end{equation}
where $r=R/R_0$ \cite{bib:Dai}. If we want to make Eq. (\ref{eq:fNKG})  and Eq. (\ref{eq:fLins}) effectively equals, $s$ should satisfy:
\begin{equation}
(s-2)+\frac{(s-4.5)r}{(1+r)}=-\frac{(\alpha + r\eta)}{(1+r)}. 
\end{equation}
Therefore, the effective lateral age parameter varies with distance $r$ as: 
\begin{equation}\label{eq:S1}
 s(r)=\frac{2-\alpha + (6.5 - \eta)r}{(1+2r)}.
\end{equation}
A similar deduction can be applied using a different class of NKG function instead of the Linsley's function. In the Ref. \cite{ropt-watson} this NKG is defined as:
\begin{equation}\label{eq:fNKGW}
  \rho (R)=k\left(\frac{R}{R_s}\right)^{-\beta}\left(1+\frac{R}{R_s}\right)^{-\beta},
\end{equation}
where $k$ is the normalization constant and $R_s=5R_0$ is the scaling parameter. From the Eq. (\ref{eq:fNKGW}) we obtain:
\begin{equation}\label{eq:dNKGW}
 \frac{d\ln\rho(r)}{d\ln r}=-\beta-\frac{\beta\left(\frac{r}{5}\right)}{(1+\frac{r}{5})},
\end{equation}
with $r=R/R_0$ again. Following the same steps as shown above, that is equaling Eqs. (\ref{eq:dNKG}) and (\ref{eq:dNKGW}), we get the effective lateral age parameter: 
\begin{equation}\label{eq:S2}
 s(r)=\frac{2-\beta + (6.9-1.4\beta)r + (1.3 - 0.4\beta)r^2}{(1+2.2r + 0.4 r^2)}.
\end{equation}
\begin{figure}[!htp]
  \centering
  \includegraphics[width=0.5\textwidth]{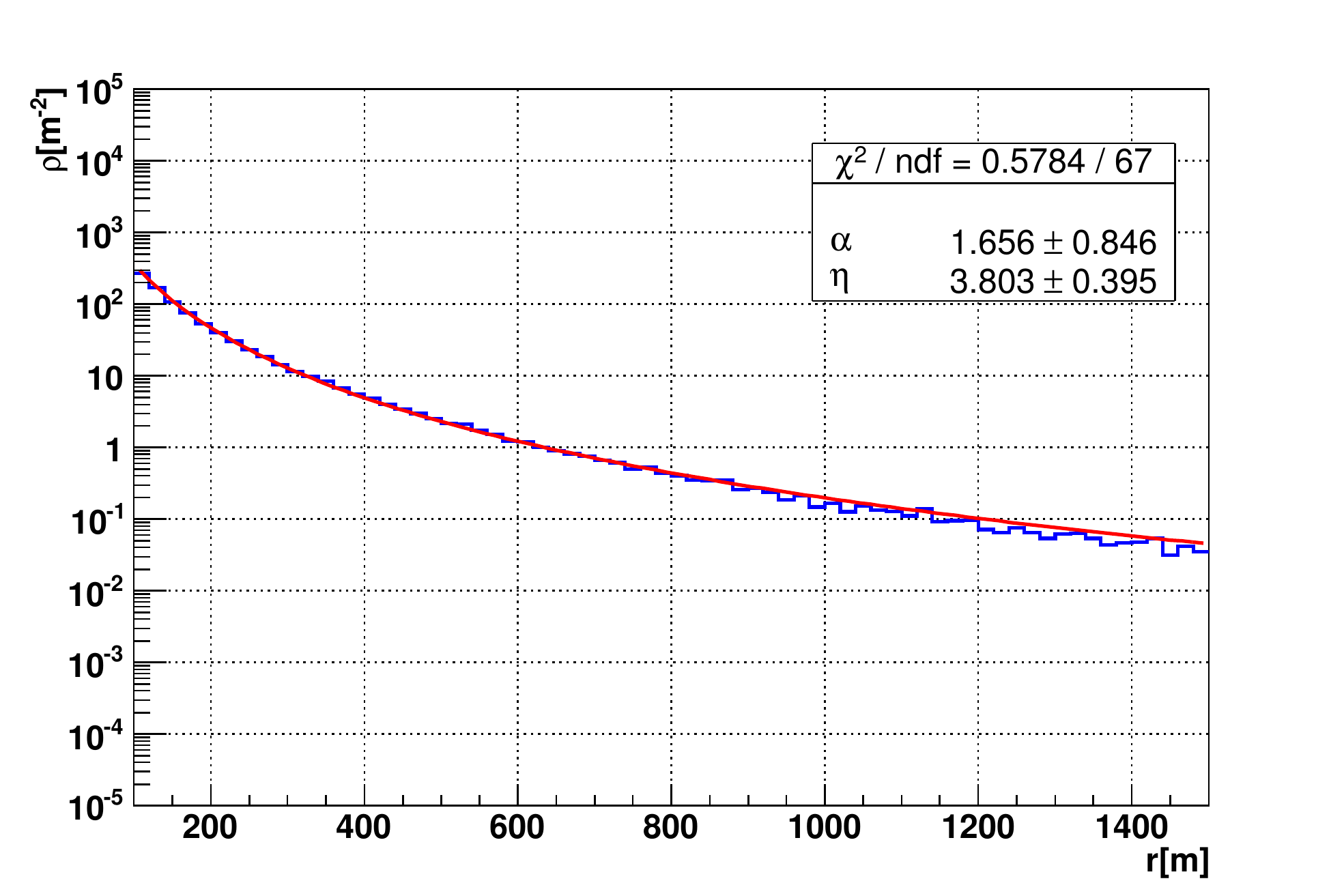}
  \caption{Lateral density distribution with iron as primary particle, energy of $10^{17.50}$ eV and zenith angle of 15$^{\circ}$. The fit was performed with Linsley's function of the Eq. (\ref{eq:fLins}).}
  \label{fit_Linsley}
\end{figure}

In order to obtain the value of the lateral shower age parameter, one can use the Eq. (\ref{eq:S1}) or the Eq. (\ref{eq:S2}), but that implies finding the parameters: $\alpha$ and $\eta$ for Eq. (\ref{eq:S1}) or $\beta$ for Eq. (\ref{eq:S2}), respectively. These parameters can be found fitting the lateral density distribution of particles on the ground, with the respective LDF function.\\

In Fig. \ref{fit_Linsley} we show the fit of the lateral density distribution corresponding to iron as primary particle, with an energy of $10^{17.50}$ eV and zenith angle of 15$^{\circ}$. The figure indicates the fit with the Linsley's function. In this case we obtained the values $\alpha=1.46$ and $\eta=3.91$. With these values and choosing one value for the distance $r$, one can find the lateral age from the Eq. (\ref{eq:S1}).\\

\begin{figure}[!htp]
  \centering
  \includegraphics[width=0.5\textwidth]{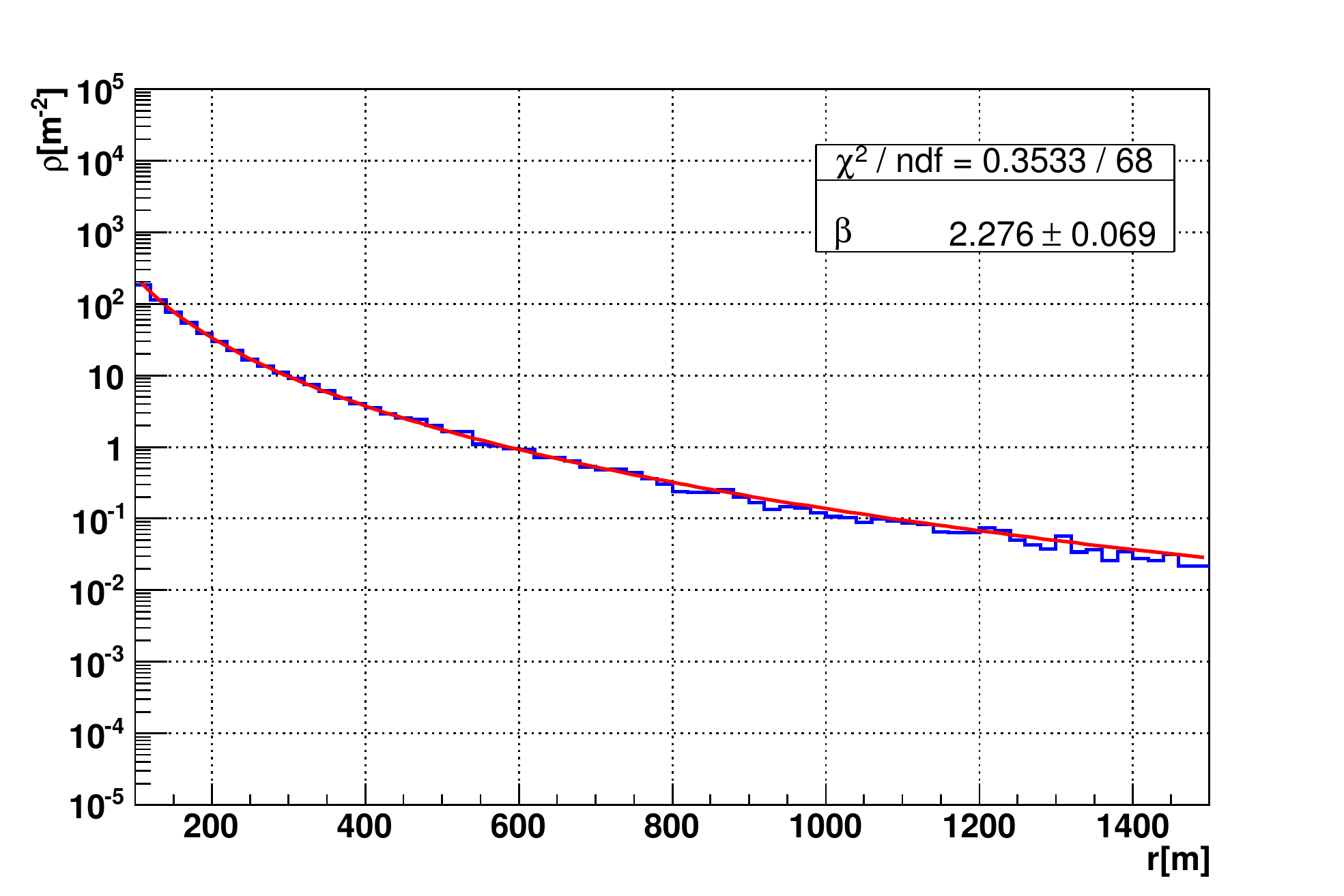}
  \caption{Lateral density distribution with proton as primary particle, energy of $10^{17.50}$ eV and zenith angle of 15$^{\circ}$. The fit was performed with NKG function of the Eq. (\ref{eq:fNKGW}).}
  \label{fit_NKG}
\end{figure}

The fit of the lateral density distribution corresponding to proton as primary particle, with an energy of $10^{17.50}$ eV and zenith angle of 15$^{\circ}$ is shown in Fig. \ref{fit_NKG}. In this case, the fit was performed with the NKG function given by the Eq. (\ref{eq:fNKGW}) and the value of $\beta$ obtained is $2.1$. Again, with that value and choosing one value for the distance $r$, one can find the lateral age from the Eq. (\ref{eq:S2}).
\begin{figure}[!htp]
  \centering
  \includegraphics[width=0.5\textwidth]{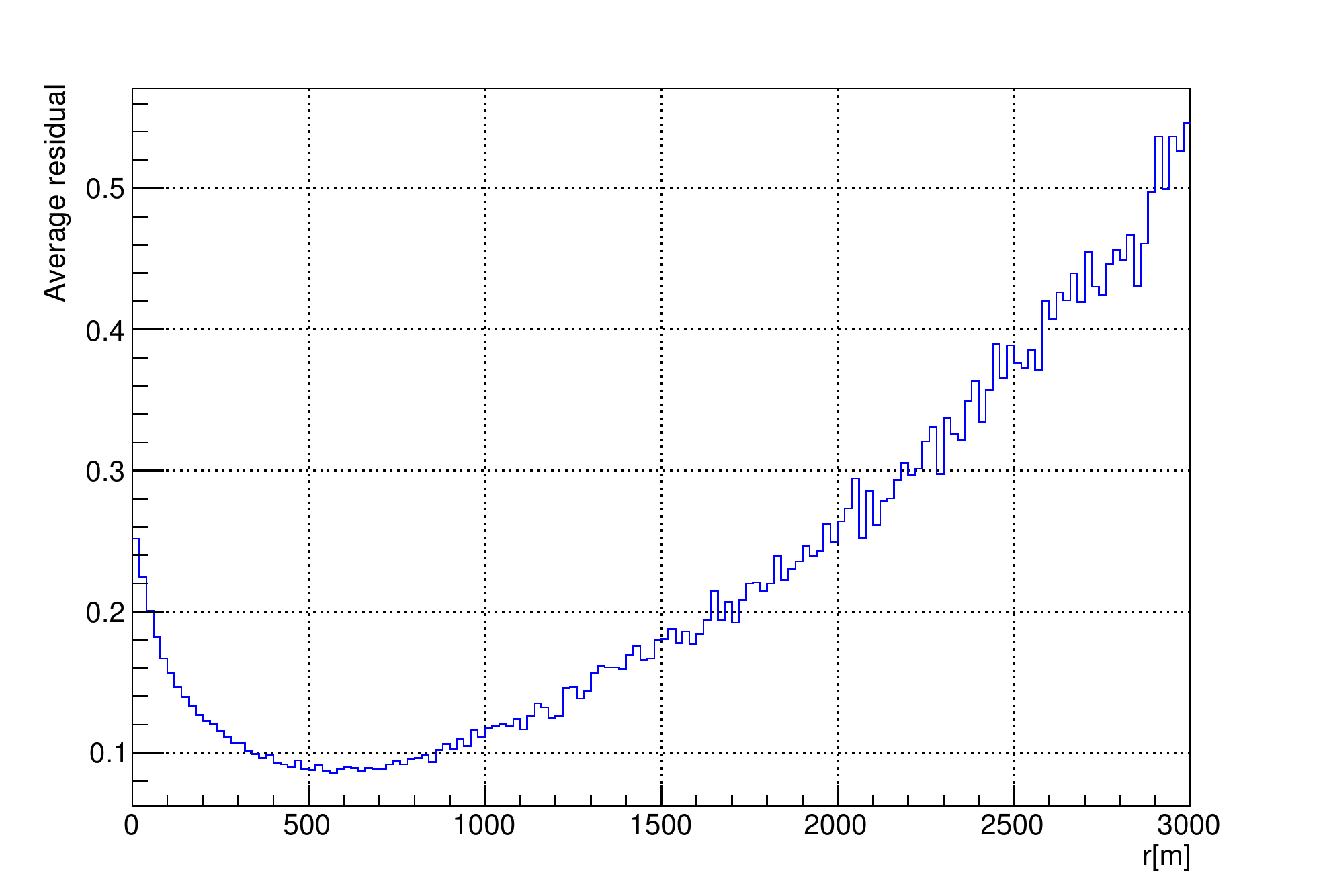}
  \caption{Average relative density residuals of 600 showers with proton as primary particle, energy of $10^{17.50}$ eV and zenith angle of 30$^{\circ}$.}
  \label{hist_ropt}
 \end{figure}
\section{Merit factor and chemical composition}
In order to study the composition discrimination power of the LSAP, we need to evaluate it at a fixed distance from the shower core. Therefore,  it is necessary to choose a distance in which the relative lateral density fluctuation is minimum \cite{ropt-watson, Hillas:00}. We will choose a distance of $500$ m. In Fig. \ref{hist_ropt} we show that $\sim 500$ m is the distance with the least fluctuation.\\ 

\begin{figure}[!htp]
  \centering
  \includegraphics[width=0.5\textwidth]{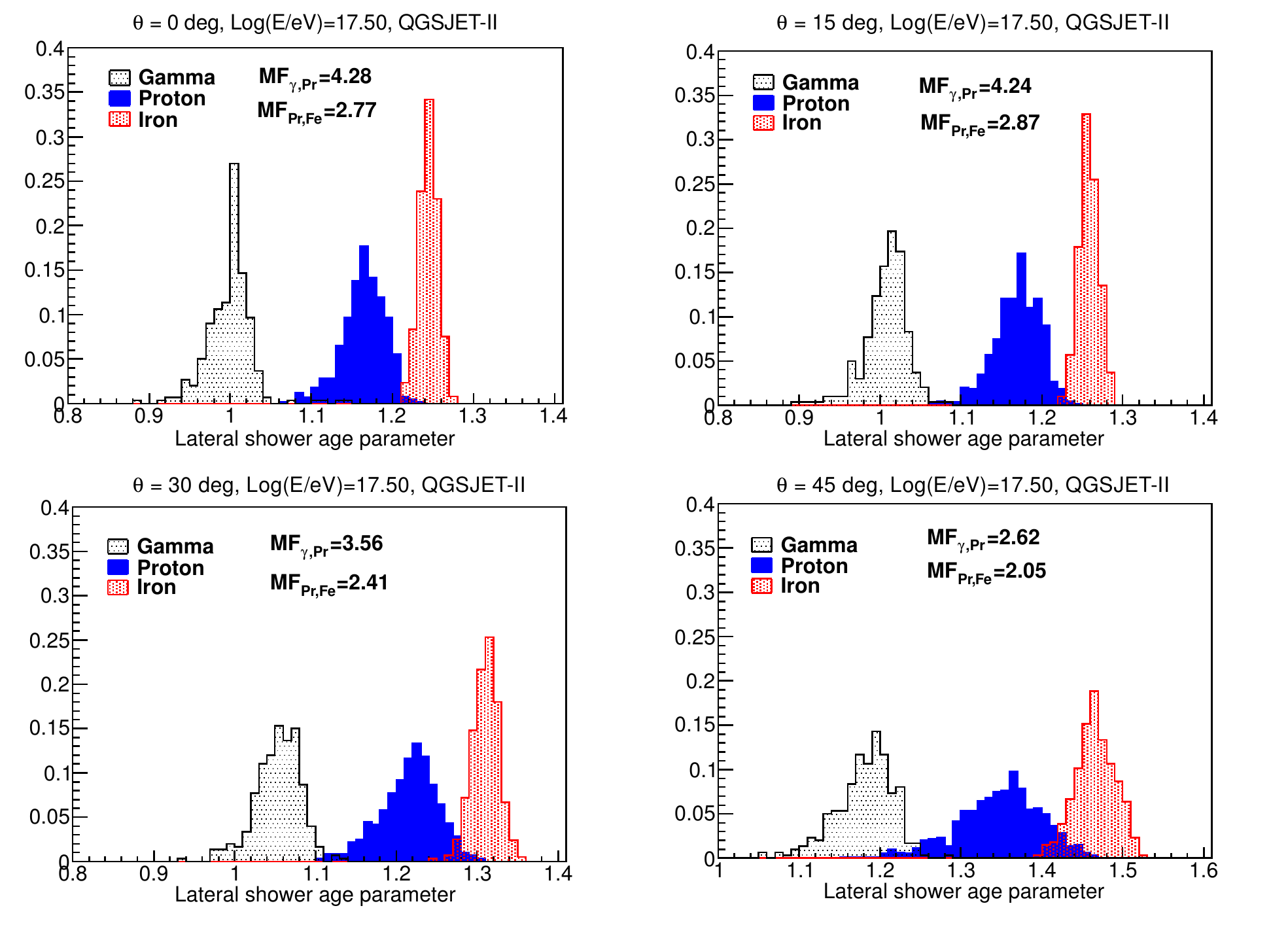}
  \caption{Lateral shower age parameter distributions using Linsley's function (see Eq. (\ref{eq:S1})). With gamma (in black), proton (in blue) and iron (in red) as primary particles, energy of $10^{17.50}$ eV and zenith angles of 0$^{\circ}$, 15$^{\circ}$, 30$^{\circ}$ and 45$^{\circ}$.}
  \label{simp_hist_lins}
 \end{figure}

\begin{figure}[!htp]
  \centering
  \includegraphics[width=0.5\textwidth]{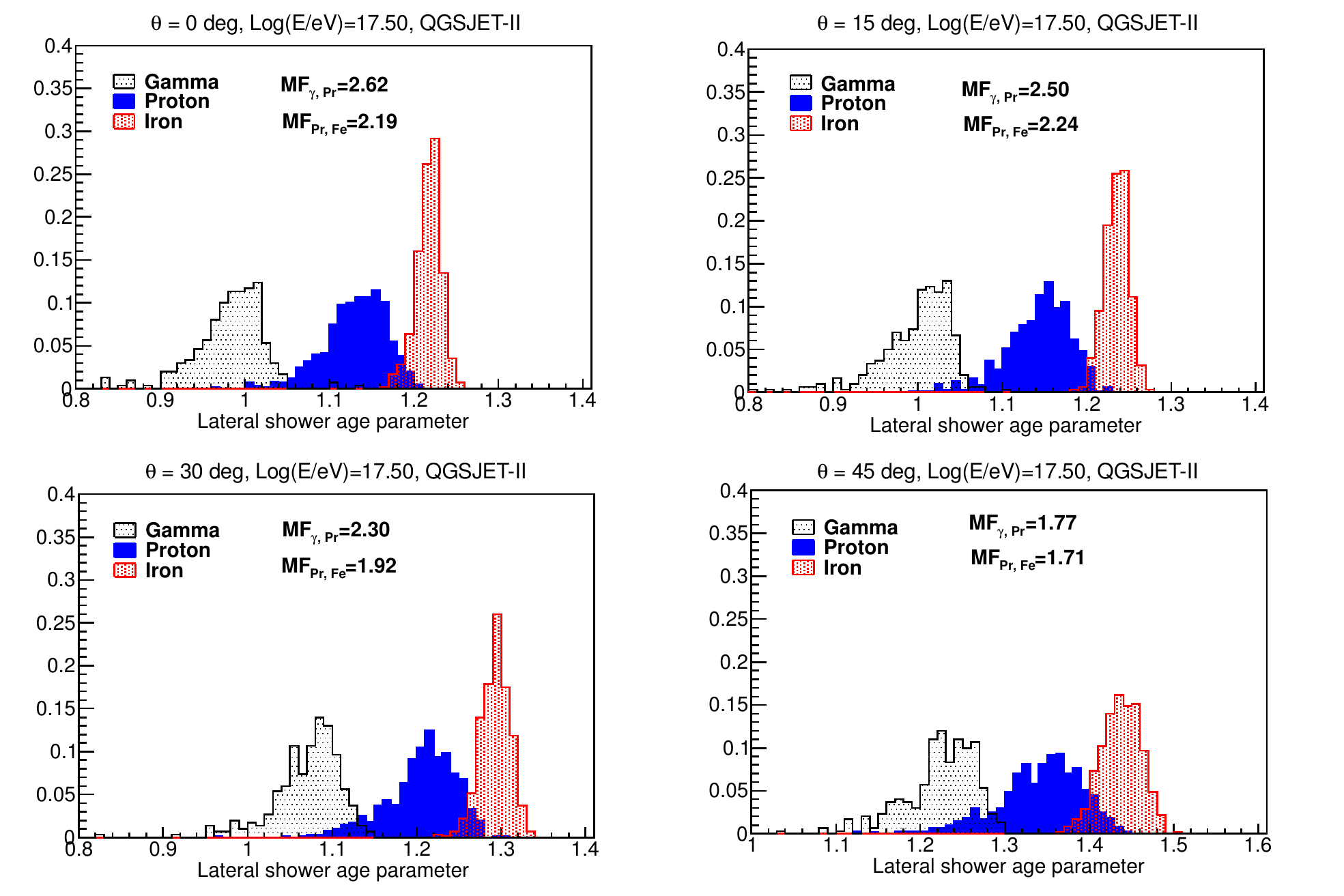}
  \caption{Lateral shower age parameter distributions using NKG function (see Eq. (\ref{eq:S2})). With gamma (in black), proton (in blue) and iron (in red) as primary particles, energy of $10^{17.50}$ eV and zenith angles of 0$^{\circ}$, 15$^{\circ}$, 30$^{\circ}$ and 45$^{\circ}$.}
  \label{simp_hist}
 \end{figure}

Once we have established the better distance to evaluate the LSAP, using  Eq. (\ref{eq:S1}) and Eq. (\ref{eq:S2}), we realize the distribution of this parameter for each primary particle, energy and zenith angle. Figures \ref{simp_hist_lins} and \ref{simp_hist} show the distributions of lateral age parameter of 600 showers, with primary energy of $10^{17.50}$ eV, four zenith angles and with proton, iron and gamma as particle primaries.\\

For each zenith angle, we calculate the merit factor between the distributions, given by:
\begin{equation}
 MF=\frac{|\langle A\rangle - \langle B\rangle|}{\sqrt{\sigma_A^2+\sigma_B^2}},
\end{equation}
where $A$ and $B$ are two distributions, respectively. From Fig. \ref{simp_hist_lins} and \ref{simp_hist} one can see that the merit factor does not depend significantly on the LDF used, but it does depend on the zenith angle (higher zenith angles correspond to lower merit factors).  This means that, when the zenith angle increases, the difficulty to determine the chemical composition of the primary particle also increases.
 
\begin{figure}[!htp]
  \centering
  \includegraphics[width=0.5\textwidth]{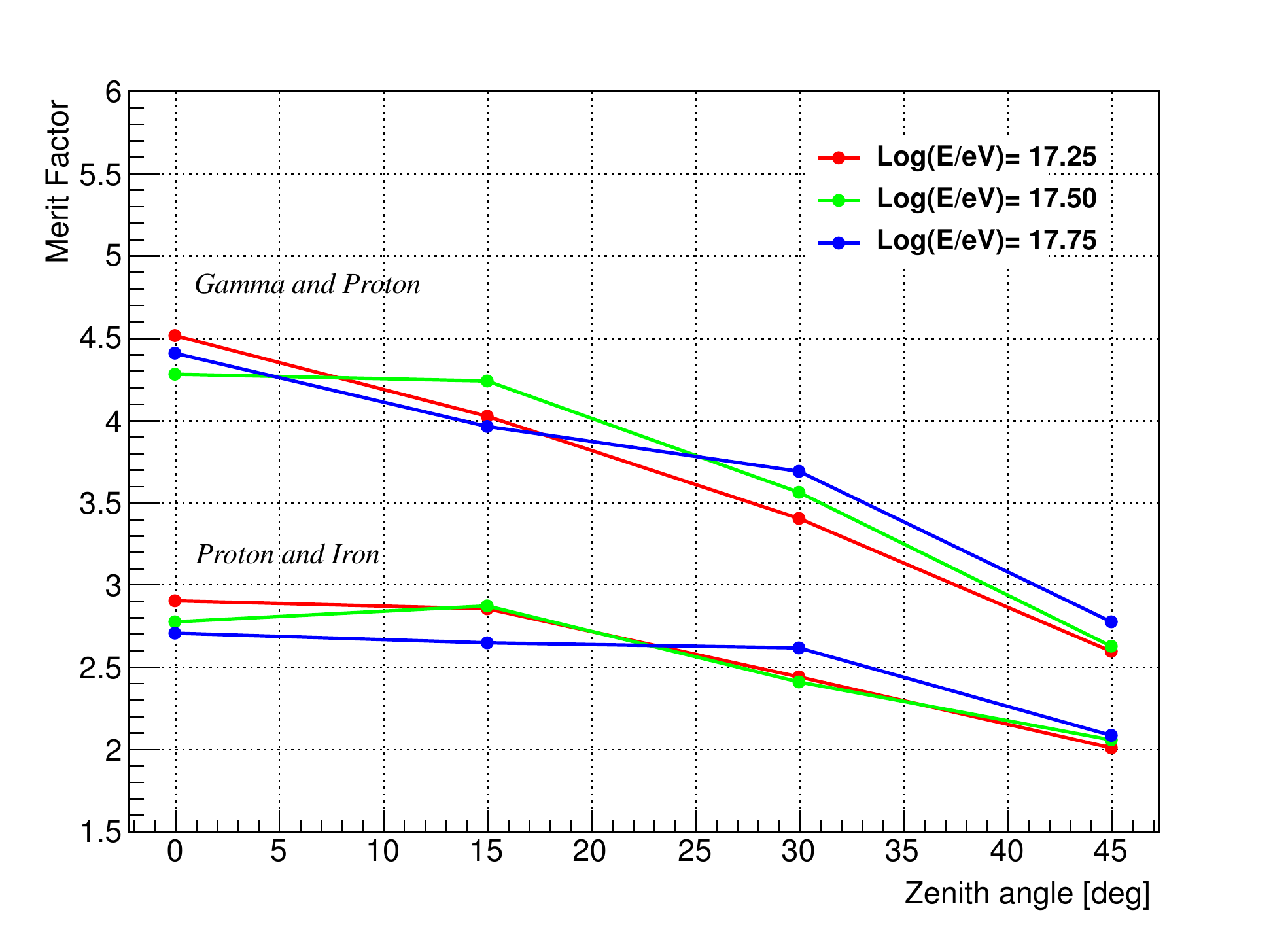}
  \caption{Merit factor as function of primary energy and zenith angle using Linsley's function.}
  \label{simp_merit_factor}
 \end{figure}

Finally, in Fig. \ref{simp_merit_factor} the merit factor as function of zenith angle is shown for three energies under consideration in this work. It can be seen that the discrimination power between gamma and nuclei is much stronger than between proton and iron but in any case MF $>2$ for all combinations studied. It is also apparent from Fig. \ref{simp_merit_factor} that merit factor is independent of the energy.  
\section{Conclusions}
In this work we performed a study of the LSAP using EAS simulated at three energies ($10^{17.25}$ eV, $10^{17.50}$ eV and $10^{17.75}$ eV). Independently of the LDF fitting function used, one can see that the LSAP distributions show a nice separation between light and heavy component. For this reason, the LSAP could be used to obtain a first estimation of the chemical composition of the primary particle. We observed that for all cases, the merit factor decreases when the zenith angle increases, independently of the primary particle energy. Therefore, for inclined showers is more difficult to discriminate between proton and iron using the LSAP.\\
Simulations of events with a ground array using water Cherenkov stations will be performed to determine how its response affects the discrimination capability of primaries considering the LSAP as estimator.

\end{document}